\documentclass{elsart3p}

\usepackage{graphicx}
 \usepackage{epsfig}

\usepackage{amssymb}
\usepackage{amsfonts}
\usepackage{bm} 

 \usepackage{natbib}

\newcommand{\be}{\begin{equation}}
\newcommand{\ee}{\end{equation}}
\newcommand{\w}[1]{\bm{#1}}
\newcommand{\volS}{\w{\epsilon}_{\scriptscriptstyle\mathcal{S}}}
\newcommand{\DS}{\w{\mathcal{D}}}
\newcommand{\el}{\w{\ell}}
\newcommand{\Sp}{{\mathcal S}}
\newcommand{\Lie}[1]{\bm{\mathcal L}_{\w{#1}}\,}
\newcommand{\LieS}[1]{{}^{\Sp}\!\Lie{#1}}

\begin{document}

\begin{frontmatter}



\title{New theoretical approaches to black holes}


\author[labluth]{Eric Gourgoulhon}
\ead{eric.gourgoulhon@obspm.fr},
\author[labiaa,labluth]{Jos\'e Luis Jaramillo}
\ead{jarama@iaa.es}

\address[labluth]{Laboratoire Univers et  Th\'eories, \\
CNRS / Observatoire de Paris / Universit\'e Paris Diderot,
92190 Meudon, France}
\address[labiaa]{Instituto de Astrof\'{\i}sica de Andaluc\'{\i}a, \\
CSIC, Apartado Postal 3004, Granada 18080, Spain}

\begin{abstract}
Quite recently, some new mathematical approaches to black holes have appeared in the literature.
They do not rely on the classical concept of event horizon --- which is very global,
but on the local concept of hypersurfaces foliated by trapped surfaces.
After a brief introduction to these new horizons, we focus on a viscous fluid analogy that can be
developed to describe their dynamics, in a fashion similar to the membrane paradigm
introduced for event horizons in the seventies, but with a significant change of
sign of the bulk viscosity.
\end{abstract}

\begin{keyword}
black hole \sep event horizon \sep trapping horizon \sep dynamical horizon \sep fluid analogy

\PACS 04.70.Bw \sep 04.20.-q \sep 04.70.-s \sep 04.25.Dm
\end{keyword}
\end{frontmatter}

\section{Introduction}

\subsection{What is a black hole ?}

\begin{figure}
\begin{center}
\includegraphics*[width=4.5cm]{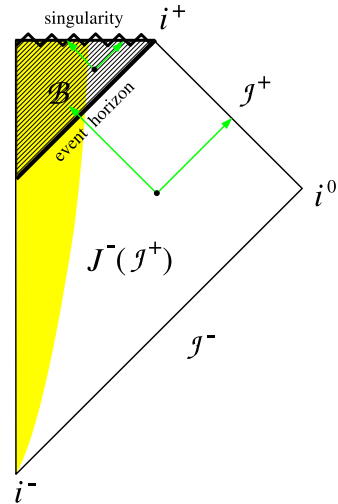}
\end{center}
\caption{Carter-Penrose diagram of a black hole $\mathcal{B}$ (cross-hatched region)
formed  by gravitational collapse of a star (colored region).
In this conformal diagram, light rays appear as straight lines inclined at $\pm45^\circ$.}
\label{f:grav_collapse}
\end{figure}

The standard mathematical definition of a \emph{black hole} is \citep{HawkiE73}
\be \label{e:def_BH}
	\mathcal{B} := \mathcal{M} - J^-(\mathcal{I}^+) ,
\ee
where $\mathcal{M}$ is a 4-dimensional manifold, endowed with a Lorentzian
metric $\w{g}$ such that $(\mathcal{M},\w{g})$ is asymptotically flat,
$\mathcal{I}^+$ is the future null infinity,
and $J^-(\mathcal{I}^+)$ is the causal past of $\mathcal{I}^+$
(cf. Fig.~\ref{f:grav_collapse}).
In common language, this means that a black hole is the
the region of spacetime where light rays cannot escape to
infinity. The \emph{event horizon} $\mathcal{H}$ is then defined
as the boundary of $\mathcal{B}$. Provided that it is smooth,
it is well known that $\mathcal{H}$ is a null hypersurface (hence it
is appears as a line inclined at $45^\circ$ in Fig.~\ref{f:grav_collapse}).

\subsection{Drawbacks of the classical definition}

As noticed by Jean-Pierre Lasota and Marek Demia\'nski long time ago \citep{DemiaL73},
definition (\ref{e:def_BH}) is not applicable is cosmology,
for usually a cosmological spacetime $(\mathcal{M},\w{g})$ is not asymptotically flat.

Moreover, even when applicable, definition (\ref{e:def_BH}) is highly non-local:
the determination of $J^-(\mathcal{I}^+)$ requires the
knowledge of the entire future null infinity.
In addition this definition has no direct relation with the notion of strong gravitational
field: as shown by \citet{AshteK04} and \citet{Krish08} on an example based on
the Vaidya metric, an event horizon can form in a flat region of
spacetime, where by \emph{flat} it is meant a vanishing Riemann tensor, i.e. no
gravitational field at all. This means that no local physical experiment whatsoever
can locate an event horizon.

Another non-local feature of event horizons is their \emph{teleological} nature
\citep{HawkiH72,Damou79,ThornPM86}.
The classical black hole boundary, i.e. the event horizon,
responds in advance to what will happen in the future. This is shown
by \citet{Booth05} on the explicit example of a black hole formed
by the collapse of two successive matter shells: after the first shell
has collapsed to form the event horizon, the latter remains stationary for
a while and then starts to grow \emph{before} the
second collapsing shell reaches it, as if it was anticipating its venue.

If one would like to deal with black holes as ``ordinary''
physical objects, like for instance in quantum gravity or numerical relativity,
the non-local (both in space and time) behavior of the event horizon mentioned above would be problematic. This has motivated the search for local characterizations of black holes.

\section{New approaches to black holes}

\subsection{Local characterizations of black holes}

The local definitions of black holes can be traced back to
the ``perfect horizons'' of \citet{Hajic73}.
However this applied only to equilibrium black holes. More recently,
the local approach has been extended to black holes out of equilibrium,
with the introduction of
\begin{itemize}
\item \emph{trapping horizons} by \citet{Haywa94},
\item \emph{isolated horizons} by \citet{AshteBF99}
\item \emph{dynamical horizons} by \citet{AshteK02},
\item \emph{slowly evolving horizons} by \citet{BoothF04}.
\end{itemize}
All these horizons are 3-dimensional submanifolds, as the event horizon.
But contrary to the latter, they rely on \emph{local} concepts.
More precisely they are all based on the notion
of \emph{trapped surfaces}, which we examine now.

\begin{figure}
\begin{center}
\includegraphics*[width=7cm]{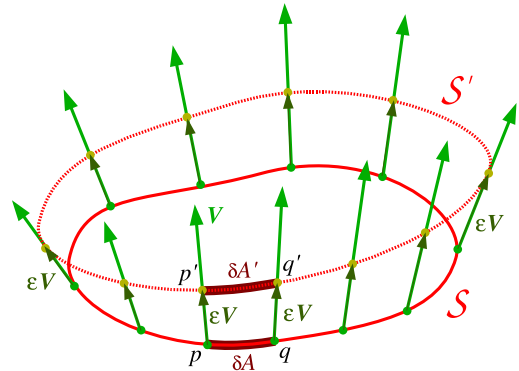}
\end{center}
\caption{Lie dragging of a spacelike 2-surface $\mathcal{S}$ along a normal
vector field $\w{V}$. In this plot, $\mathcal{S}$ appears as a closed line, whereas
it is actually 2-dimensional.}
\label{f:expansion6}
\end{figure}

\subsection{Trapped surface}

Before defining a trapped surface, let us start by the general concept
of the expansion of a surface along a normal vector field.
Consider a spacelike 2-surface $\mathcal{S}$, as in
Fig.~\ref{f:expansion6}.
Take a vector field $\w{V}$ defined on $\mathcal{S}$ and normal to $\mathcal{S}$
at each point. For a given small parameter $\varepsilon\in\mathbb{R}$,
displace the point $p\in\mathcal{S}$
 by the vector $\varepsilon \w{V}$ to the point $p'$.
Repeat this for each point in $\mathcal{S}$, keeping the value of $\varepsilon$
fixed. This defines a new surface $\mathcal{S}'$. This process is called \emph{Lie dragging}
along the vector field $\w{V}$.
At each point, the \emph{expansion of $\mathcal{S}$ along $\w{V}$} is defined from the relative change in the area element $\delta A$ around that point (cf. Fig.~\ref{f:expansion6}):
\be \label{e:def_expansion}
	\theta^{(\w{V})} := \lim_{\varepsilon\rightarrow 0}
	\frac{1}{\varepsilon}\frac{\delta A'-\delta A}{\delta A}
  = \mathcal{L}_{\w{V}} \ln \sqrt{q} = q^{\mu\nu} \nabla_\mu V_\nu ,
\ee
where $\w{q}$ denotes the metric induced on $\mathcal{S}$ by the spacetime
metric $\w{g}$, $q$ the determinant of $q_{\mu\nu}$, $\mathcal{L}_{\w{V}}$
the Lie derivative along the vector field $\w{V}$ and $\nabla$ the spacetime
covariant derivative.

\begin{figure}
\begin{center}
\includegraphics*[width=7cm]{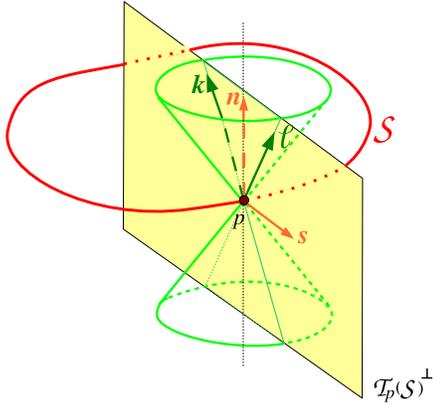}
\end{center}
\caption{Null directions $\w{\ell}$ and $\w{k}$ normal to a closed spacelike
2-surface $\mathcal{S}$. As in Fig.~\ref{f:expansion6}, $\mathcal{S}$ is drawn
as a 1-dimensional contour, instead of a 2-dimensional surface.
$\mathcal{T}_p(\mathcal{S})^\perp$ is the 2-plane normal to $\mathcal{S}$ at the point
$p$. Its intersection with the light cone emanating from $p$ defines the null directions
$\w{\ell}$ and $\w{k}$. The unit vectors $\w{n}$ and
$\w{s}$ are respectively a timelike normal and a spacelike normal to $\mathcal{S}$.}
\label{f:surf_cone3}
\end{figure}

With this definition of the expansion in hand, we are ready to define a
trapped surface as follows. Consider a \emph{closed} (i.e. compact without boundary)
and \emph{spacelike} 2-dimensional surface $\mathcal{S}$
embedded in the spacetime $(\mathcal{M},\w{g})$.
Being spacelike, $\mathcal{S}$ lies outside the light cone (cf. Fig.~\ref{f:surf_cone3}),
which means that there exist two future-directed null directions
orthogonal to $\mathcal{S}$:
$\w{\ell}$, the so-called \emph{outgoing null normal},
and $\w{k}$, the so-called \emph{ingoing null normal}.
Note that $\w{\ell}$ and $\w{k}$ are defined up to a rescaling:
$\w{\ell'} = \lambda \w{\ell}$ and $\w{k'} = \mu \w{k}$.

In flat space, the expansion of $\mathcal{S}$ along $\w{\ell}$ is always positive:
$\theta^{(\w{\ell})} >0$, whereas that along $\w{k}$ is negative : $\theta^{(\w{k})}<0$.
Now the surface $\mathcal{S}$ is called \emph{trapped}
iff both expansions are negative: $\theta^{(\w{\ell})} < 0$ and $\theta^{(\w{k})}<0$.
The limiting case, $\theta^{(\w{\ell})} = 0$ and $\theta^{(\w{k})}<0$, is called
a \emph{marginally trapped surface}. These definitions have been introduced by \citet{Penro65}. They clearly constitute a local concept\footnote{some
authors say \emph{quasilocal} instead of local, because the definition relies on the
notion of a surface, and not merely a point}. Moreover, this concept is related to very
strong gravitational fields, since for weak fields, one has clearly $\theta^{(\w{\ell})} >0$.

It is worth noticing that in the previously mentioned work by \citet{DemiaL73}, 
the ``local event horizon'' defined by the authors
is nothing but a marginally trapped surface.

\subsection{Link with apparent horizons}

A closed spacelike 2-surface $\mathcal{S}$ is said to be \emph{outer trapped}
(resp. \emph{marginally outer trapped}) if, and only if, \citep{HawkiE73}
\begin{itemize}
\item the notions of \emph{interior} and \emph{exterior} of $\mathcal{S}$ can be defined
(for instance spacetime asymptotically flat);
\item the \emph{outgoing} null normal $\w{\ell}$ satisfies
$\theta^{(\w{\ell})} < 0$ (resp. $\theta^{(\w{\ell})} =0$).
\end{itemize}
Notice that no condition is imposed
on the expansion $\theta^{(\w{k})}$ along the ingoing null normal.

\begin{figure}
\begin{center}
\includegraphics*[width=8cm]{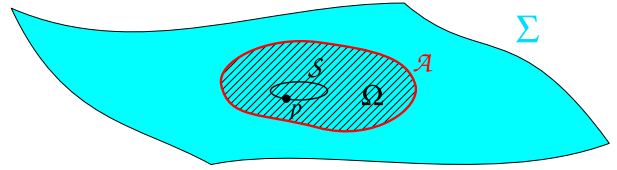}
\end{center}
\caption{Spacelike hypersurface $\Sigma$ containing an outer trapped region $\Omega$
(i.e. a set of points through which there is at least one outer trapped surface).
The apparent horizon $\mathcal{A}$ is then the boundary of $\Omega$.}
\label{f:app_horizon}
\end{figure}

Let us then consider a spacelike hypersurface $\Sigma$ extending to spatial infinity
(Cauchy surface) (cf. Fig.~\ref{f:app_horizon}). The \emph{outer trapped region} of $\Sigma$
is defined as the set $\Omega$ of points
$p\in\Sigma$  through which there is a outer trapped surface $\mathcal{S}$
lying in $\Sigma$. An \emph{apparent horizon} in $\Sigma$ is
then a connected component $\mathcal{A}$ of the boundary of $\Omega$ \citep{HawkiE73}. Then a classical result by \citet{HawkiE73} states that the apparent horizon is a marginally
outer trapped surface \citep[see also the recent study by][]{AnderM07}.

\subsection{Connection with singularities and black holes}

A famous theorem by \citet{Penro65} makes the link with the trapped surfaces
introduced above and spacetime singularities: provided that the weak energy condition holds,
if there exists a trapped surface $\mathcal{S}$, then there exists a
singularity in $(\mathcal{M},\w{g})$ (in the form of a future inextendible
null geodesic).
Another theorem by \citet{HawkiE73} states that, provided that the cosmic censorship conjecture holds, if the spacetime contains a trapped surface $\mathcal{S}$,
then it necessarily contains
a black hole $\mathcal{B}$ and $\mathcal{S}\subset\mathcal{B}$.

\subsection{Local definitions of ``black holes''}

\begin{figure}
\begin{center}
\includegraphics*[width=5cm]{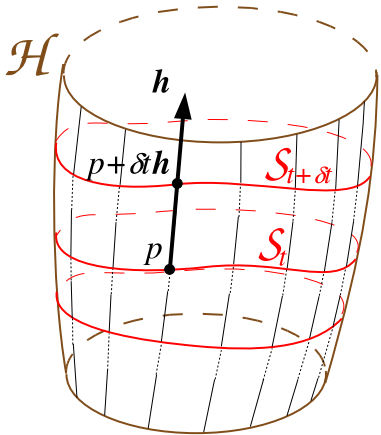}
\end{center}
\caption{Hypersurface $\mathcal{H}$ foliated by a 1-parameter family of 2-surfaces
$(\mathcal{S}_t)_{t\in\mathbb{R}}$. $\w{h}$ is the canonical evolution vector
associated with the parameter $t$.}
\label{f:hyp_foliat_h}
\end{figure}

Having recalled the previous classical results about trapped surfaces and
apparent horizons, we now state the local definitions of black hole horizons,
alternative to the event
horizon, that have appeared quite recently in the literature.
A hypersurface $\mathcal{H}$ of $(\mathcal{M},\w{g})$ is said to be
\begin{itemize}

\item a \emph{future outer trapping horizon (FOTH)} iff
(i) $\mathcal{H}$ foliated by marginally trapped 2-surfaces :
$\mathcal{H}=\bigcup_{t\in\mathbb{R}} \mathcal{S}_t$ with
$\theta^{(\w{k})} < 0$
and  $\theta^{(\w{\ell})} = 0$ (cf. Fig.~\ref{f:hyp_foliat_h}), and (ii)
the outermost condition $\mathcal{L}_{\w{k}}\theta^{(\w{\ell})}<0$ is satisfied
\citep{Haywa94};

\item a \emph{dynamical horizon (DH)} iff
(i) $\mathcal{H}$ is foliated by marginally trapped 2-surfaces and
(ii) $\mathcal{H}$ is spacelike \citep{AshteK02};

\item a \emph{non-expanding horizon\footnote{the non-expanding horizons were called
\emph{perfect horizons} by  \citet{Hajic73}}} iff
(i) $\mathcal{H}$ is a null hypersurface (with null normal $\w{\ell}$ say)
and  (ii) $\theta^{(\w{\ell})} = 0$ \citep{Hajic73};

\item an \emph{isolated horizon} iff
(i) $\mathcal{H}$ is a non-expanding horizon (it has then a well defined
geometry, with a unique connection $\hat{\w{\nabla}}$, despite the induced
metric is degenerate)
and (ii) $\mathcal{H}$'s geometry is not evolving
along the null generators: $[\mathcal{L}_{\w{\ell}}, \hat{\w{\nabla}}]=0$
\citep{AshteBF99}.
\end{itemize}

Note that in generic dynamical situations, the notions
of FOTH and DH are equivalent \citep{Booth05}. In stationary
situations, a FOTH becomes a null hypersurface, whereas a DH (which by definition is spacelike) cannot exist; it should be replaced by the notion of isolated horizon \citep{AshteBF99,AshteK04,Booth05,GourgJ06a}.
If $\mathcal{H}$ is an event horizon, the 2-surfaces $\mathcal{S}_t$ are not marginally trapped, except in stationary configurations (Kerr black hole). On the contrary
they are expanding, by the famous \citet{Hawki72} area increase law :
$\theta^{(\w{\ell})}>0$.

These ``new'' horizons have their own dynamics,
ruled by Einstein equations.
In particular, one can establish for them
existence and (partial) uniqueness theorems \citep{AnderMS05,AshteG05},
first and second laws analogous to the classical laws of black hole mechanics
\citep{AshteK03,Haywa04b},
a viscous fluid bubble analogy (``membrane paradigm'' as for the event horizon),
leading to a Navier-Stokes-like equation
\citep{Gourg05,GourgJ06b}. Recent review articles on the subject
are \citet{AshteK04,Booth05,GourgJ06a,Krish08}.
Notice that the FOTH and DH proved to be useful in
numerical relativity \citep{SchneKB06,JaramAL07,JaramGCI08,JaramVG08}
not only in the ``follow-up'' of the horizon, but also
in the prescription of excised black hole initial data in
quasi-equilibrium \citep{JarGouMen04,DaiJarKri05,CooPfe04,Ans05,CauCooGri06}.

\section{Geometry of hypersurface foliations by spacelike 2-surfaces}

Since the trapping and dynamical horizons are based on a foliation $(\mathcal{S}_t)_{t\in\mathbb{R}}$ by closed spacelike 2-surfaces
of a hypersurface $\mathcal{H}$, let
us first discuss the geometrical properties of such foliations.

\subsection{Relevant vectors} \label{s:vectors}

\begin{figure}
\begin{center}
\includegraphics*[width=7cm]{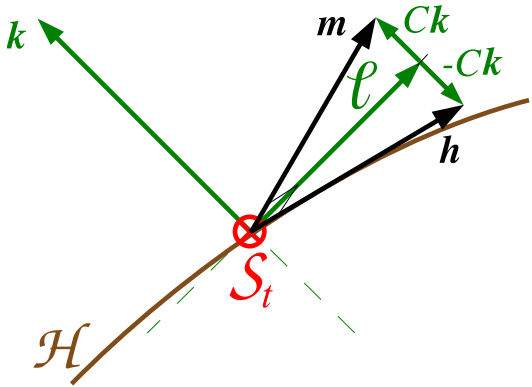}
\end{center}
\caption{Evolution vector $\w{h}$, null normals $\w{\ell}$ and $\w{k}$ and
normal vector $\w{m}$ along a foliated hypersurface $\mathcal{H}$.
The figure is drawn in the plane normal to $\mathcal{S}_t$, which is reduced to a point,
$\mathcal{H}$ being reduced to a line.}
\label{f:coupe_normal_m}
\end{figure}

We shall call \emph{evolution vector} the unique vector field
$\w{h}$ that is tangent to
$\mathcal{H}$, orthogonal to $\mathcal{S}_t$ and such that $\mathcal{L}_{\w{h}}t=1$, where $\mathcal{L}_{\w{h}}$ denotes
the Lie derivative along $\w{h}$: $\mathcal{L}_{\w{h}}t=h^\mu\partial_\mu t$
(cf. Fig.~\ref{f:hyp_foliat_h}).
The latter property implies that the 2-surfaces $\mathcal{S}_t$ are Lie dragged
to each other by $\w{h}$.
Let $C$ be half the scalar square of $\w{h}$ with respect to the metric $\w{g}$:
\be
	\w{h}\cdot\w{h}=2C
\ee
(we systematically denote the scalar product corresponding to the spacetime metric $\w{g}$ with
a dot).
It is easy to see that the sign of $C$ gives the signature
of the hypersurface $\mathcal{H}$: $C$ is positive, zero and negative for respectively
spacelike, null and timelike hypersurfaces.
There exists a unique pair $(\w{\ell},\w{k})$ of null vectors normal to $\mathcal{S}_t$ and
a unique vector $\w{m}$ normal to $\mathcal{H}$ such that (cf. Fig.~\ref{f:coupe_normal_m})
\be \label{e:def_l_k_m}
    \w{h} = \w{\ell} - C \w{k},\quad
    \w{m} = \w{\ell} + C \w{k}\quad\mbox{and}\quad
    \w{\ell}\cdot\w{k} = - 1 .
\ee
For any vector field $\w{v}$ normal to $\mathcal{S}_t$, such as $\w{h}$, $\w{m}$, $\w{\ell}$ or $\w{k}$, we define the  \emph{shear tensor}
$\w{\sigma}^{(\w{v})}$ of the surface $\mathcal{S}_t$ when Lie-dragged along $\w{v}$
by
\be \label{e:def_exp_shear}
   \mathcal{L}_{\w{v}} \w{q} = \theta^{(\w{v})} \w{q} + 2 \w{\sigma}^{(\w{v})}
    \quad\mbox{and}\quad
     \mathrm{tr}\, \w{\sigma}^{(\w{v})} = 0 ,
\ee
where, as before, $\w{q}$ is the induced metric on $\mathcal{S}_t$ ($\w{q}$ is positive definite
since $\mathcal{S}_t$ is assumed to be spacelike) and $\mathcal{L}_{\w{v}} \w{q}$ is its Lie derivative
resulting from the dragging of the surface $\mathcal{S}_t$ along the normal vector $\w{v}$.
The vanishing of the trace of $\w{\sigma}^{(\w{v})}$
with respect to the metric $\w{q}$ is a consequence of the definition
(\ref{e:def_expansion}) of $\theta^{(\w{v})}$.

Let us denote by $\kappa$ the component along $\el$ of the ``acceleration'' of $\w{h}$ in the decomposition \citep{GourgJ06b}
\be \label{e:def_kappa}
	\w{\nabla}_{\w{h}} \w{h} = \kappa \, \el + (C\kappa - \Lie{h} C) \w{k}
	- \DS C .
\ee
If $\mathcal{H}$ is an event horizon, then it is a null hypersurface,
so that $\w{h}=\w{\ell}$, $C=0$ and the above relation reduces to
\be \label{e:def_kappa_EH}
	\w{\nabla}_{\w{\ell}} \w{\ell} = \kappa \, \w{\ell},
\ee
showing that, in this case, $\kappa$ is nothing but the surface gravity of the black hole.

\subsection{Extrinsic geometry of the 2-surfaces}

On each 2-surface $\mathcal{S}_t$, let us denote by $\DS$ the connection
compatible with the induced metric $\w{q}$ (this connection is unique since
$\w{q}$ is not degenerate). $\DS$ is related to the
spacetime connection $\w{\nabla}$ via the
the \emph{second fundamental tensor} of $\mathcal{S}_t$, $\w{\mathcal{K}}$
\citep{Carte92a,Senov05}:
\be
    \forall (\w{u},\w{v}) \in \mathcal{T}(\mathcal{S}_t)^2,\quad
    \w{\nabla}_{\w{u}} \w{v} = \DS_{\w{u}} \w{v}
        + \w{\mathcal{K}}(\w{u},\w{v}) .
\ee
$\w{\mathcal{K}}$ is a type $(1,2)$ tensor, which is expressible in
term of the covariant derivative of $\w{q}$, according to
\be	
	\mathcal{K}^\alpha_{\ \, \beta\gamma} =
        \nabla_\mu q^\alpha_{\ \, \nu} \;
        q^\mu_{\ \, \beta} q^\nu_{\ \, \gamma} .
\ee
Contrary to the case of a hypersurface\footnote{for a non-degenerate hypersurface,
$\w{\mathcal{K}}$ is related to the second fundamental \emph{form} $\w{K}$
(also called \emph{extrinsic curvature tensor}) via
$\mathcal{K}^\alpha_{\ \, \beta\gamma} = - n^\alpha K_{\beta\gamma}$,
where $\w{n}$ is the unit normal to the hypersurface},
the extrinsic geometry of the 2-surface
$\mathcal{S}_t$ is not entirely specified by
$\w{\mathcal{K}}$. The latter encodes only the part of the
variation of $\mathcal{S}_t$'s normals which is parallel to $\mathcal{S}_t$.
The remaining part, i.e. the
variation of the two normals with respect to each other, is
encoded by the \emph{normal fundamental forms}
(also called \emph{external rotation coefficients} or
\emph{connection on the normal bundle}, or if $\mathcal{H}$ is null,
\emph{H\'a\'\j i\v{c}ek 1-form}), defined by \citep[see e.g.][]{Haywa94}
\begin{eqnarray}
	 \w{\Omega}^{(\w{\ell})} & := &  -
        \w{k} \cdot \w{\nabla}_{\vec{\w{q}}}\,  \w{\ell} \label{e:def_Omega_l} \\
 \w{\Omega}^{(\w{k})} & := &  -
        \w{\ell} \cdot \w{\nabla}_{\vec{\w{q}}}\,  \w{k} ,
\end{eqnarray}
where $\vec{\w{q}}$ denotes the orthogonal projector on the surface
$\mathcal{S}_t$.
In terms of components, Eq.~(\ref{e:def_Omega_l}) is written
\be
\Omega^{(\w{\ell})}_\alpha := -
     k_\mu \nabla_\nu \ell^\mu \,  q^\nu_{\ \, \alpha} ,
\ee
with a similar relation for $\Omega^{(\w{\ell})}_\alpha$.
Thanks to the relation $\w{\ell}\cdot\w{k}=-1$ [Eq.~(\ref{e:def_l_k_m})],
we have $\w{\Omega}^{(\w{k})} = - \w{\Omega}^{(\w{\ell})}$.
Note that contrary to the second fundamental tensor $\w{\mathcal{K}}$, the
normal fundamental forms are not unique: any rescaling
$\w{\ell'} = \lambda \w{\ell}$ of the null normal
results in
\be \label{e:rescale_Omega}
	\w{\Omega}^{(\w{\ell'})} = \w{\Omega}^{(\w{\ell})} + \DS \ln \lambda .
\ee

\section{A Navier-Stokes-like equation}

\subsection{Concept of black hole viscosity}

When studying the response of the event horizon to external perturbations in the
early seventies,
\citet{HawkiH72} and \citet{Hartl73} introduced the concept of
\emph{black hole viscosity}.
This fluid analogy took its full significance when \citet{Damou79,Damou82}
derived from Einstein equation a 2-dimensional Navier-Stokes like equation
governing the evolution of the event horizon, and letting appear
some \emph{shear viscosity} and well as some \emph{bulk viscosity}.
The 2-dimensional fluid (membrane) point of view has been further developed in the
famous \emph{Membrane Paradigm} book
by \citet{ThornPM86}.

A natural question which then arises is:
shall we restrict the analysis to the event horizon ?
In other words, can we extend the concept of viscosity to the local characterizations of black
hole recently introduced, i.e. FOTH and DH ?

A priori this does not seem obvious because, from a pure geometrical point of view,
the event horizon and the ``local'' horizons are of different type:
the event horizon is always a null hypersurface (hence is endowed with
a degenerate metric), whereas a DH is always a spacelike
surface (hence with a positive definite metric) and a
FOTH can be either null or spacelike.
We shall see that nevertheless the fluid analogy can also be extended to these horizons,
with some significant change in the sign of the bulk viscosity.

\subsection{Original Damour-Navier-Stokes equation} \label{s:DNS_ori}

Damour considered the case where $\mathcal{H}$ is a black hole event horizon.
In particular it is a null hypersurface and the null vector $\w{\ell}$ is
normal to it. From the Einstein equation,
he has derived the relation \citep{Damou79,Damou82} \citep[see also][]{DamouL08}
\begin{eqnarray}
    \LieS{\el}  \w{\pi} + \theta^{(\el)} \w{\pi}
    & = & - \DS P
     + 2 \mu  \DS \cdot \vec{\w{\sigma}}^{(\el)}
    + \zeta \DS \theta^{(\el)}\nonumber \\
	& &  + \w{f} , \label{e:DNS_ori}
\end{eqnarray}
where
\begin{itemize}
\item  $\w{\pi} := -{1}/({8\pi})\;  \w{\Omega}^{(\el)} $ is analogous to some
momentum surface density,
\item $P := {\kappa}/(8\pi)$ is analogous to the  pressure
[$\kappa$ being defined by Eq.~(\ref{e:def_kappa_EH})],
\item $\mu := {1}/(16\pi)$ is analogous to the shear viscosity,
\item $\zeta := -1/(16\pi)$ is analogous to the bulk viscosity,
\item $\w{f} := - \w{T}(\el,\vec{\w{q}})$ is the external force surface density,
$\w{T}$ being the stress-energy tensor of any matter or electromagnetic field
present around the horizon.
\end{itemize}
Equation~(\ref{e:DNS_ori}) is structurally identical to a
Navier-Stokes equation for a 2-dimensional fluid.
The reader is referred to Chap.~VI of \citet{ThornPM86} for an extended
discussion of this analogy with a viscous fluid \citep[see also Sec.~2.3 of][]{DamouL08}.

A striking feature of the above Navier-Stokes equation (\ref{e:DNS_ori}) is that the bulk viscosity
is negative:
\be \label{e:zeta_EH}
	\zeta = \zeta_{\rm EH} = -\frac{1}{16\pi} < 0 ,
\ee
where the subscript EH stands for ``event horizon''.
For an ordinary fluid,
this negative value would yield to a dilation or contraction instability.
This is in agreement with the well-known tendency of a null hypersurface
to continually contract or expand. However the
event horizon is stabilized by the teleological condition that
its expansion must vanish in the far future, when an equilibrium state
has been reached \citep{Damou79}.

\subsection{Generalization to the non-null case} \label{s:gen_non-null}

In order to generalize Eq.~(\ref{e:DNS_ori}) to the case where
$\mathcal{H}$ is not necessarily a null hypersurface, it is worth
to notice that in Eq.~(\ref{e:DNS_ori}), the vector $\el$ plays two
role: it is the natural evolution vector along $\mathcal{H}$ and
it is also the normal to $\mathcal{H}$. In the non null case, these
two role are played respectively by the vectors $\w{h}$ and $\w{m}$
introduced in Sec.~\ref{s:vectors}. Of course, at the null limit
($C=0$), $\w{h} = \w{m} = \el$ [cf. Eq.~(\ref{e:def_l_k_m})].

Having realized this, the starting point of the calculation is the
contracted Ricci identity applied to the vector $\w{m}$
and projected onto $\Sp_t$:
\be \label{e:Ricci_m_q}
    \left( \nabla_\mu \nabla_\nu m^\mu
    - \nabla_\nu \nabla_\mu m^\mu \right) q^\nu_{\ \, \alpha}  =
    R_{\mu\nu} m^\mu q^\nu_{\ \, \alpha} ,
\ee
where $R_{\mu\nu}$ is the spacetime Ricci tensor, to be replaced
ultimately by its expression in terms of the matter stress-energy
tensor $T_{\mu\nu}$ according to Einstein equation.
After some manipulations, one arrives at \citep{Gourg05}
\begin{eqnarray}
& & \LieS{\w{h}} \w{\Omega}^{(\el)}
   + \theta^{(\w{h})} \, \w{\Omega}^{(\el)}
    = \DS \kappa
        - \DS \cdot \vec{\w{\sigma}}^{(\w{m})}
             + \frac{1}{2} \DS \theta^{(\w{m})}\nonumber \\
& & \qquad
      - \theta^{(\w{k})} \DS C
    + 8\pi \w{T}(\w{m},\vec{\w{q}}) . 	\label{e:DNS_gen}
\end{eqnarray}
In the null limit, $C=0$, $\w{h} = \w{m} = \el$, and the above
equation reduces to the original Damour-Navier-Stokes equation
(\ref{e:DNS_ori}). On the other side, if $\mathcal{H}$ is a
FOTH or a DH, then $\theta^{(\w{m})}=-\theta^{(\w{h})}$
(since in this case $\theta^{(\w{\ell})}=0$ and we can deduce
from Eq.~(\ref{e:def_l_k_m}) the relation
$\theta^{(\w{m})}=-\theta^{(\w{h})}+2\theta^{(\w{\ell})} $)
and Eq.~(\ref{e:DNS_gen}) can be written
\begin{eqnarray}
    \LieS{\el}  \w{\pi} + \theta^{(\w{h})} \w{\pi}
    & = & - \DS P
     + \frac{1}{8\pi}  \DS \cdot \vec{\w{\sigma}}^{(\w{m})}
    + \zeta \DS \theta^{(\w{h})}\nonumber \\
	& &  + \w{f} , \label{e:DNS_FOTH}
\end{eqnarray}
where $\w{f} := - \w{T}(\w{m},\vec{\w{q}}) + \theta^{(\w{k})}/(8\pi)\,  \DS C$
and, as in Eq.~(\ref{e:DNS_ori}), $\w{\pi} := -{1}/({8\pi})\;  \w{\Omega}^{(\el)} $,
$P:=\kappa/(8\pi)$, but contrary to Eq.~(\ref{e:DNS_ori}),
\be \label{e:zeta_FOTH}
	\zeta = \zeta_{\rm FOTH} := \frac{1}{16\pi} > 0 .
\ee
This positive value of the bulk viscosity shows that
FOTHs and DHs behave as ``ordinary''
physical objects.

\subsection{Angular momentum flux law}

In general relativity, the angular momentum is usually well defined only
if there exists a Killing vector field $\w{\varphi}$ which generates
a symmetry around some axis. To generalize the definition of angular momentum
to the cases where no symmetry is present, let us follow \citet{BoothF05} and introduce a vector field on $\mathcal{H}$ which
\begin{itemize}
\item is tangent to $\Sp_t$
\item has closed orbits
\item has vanishing divergence with respect to the induced metric on $\Sp_t$:
\be	\label{e:divvp}
	\DS\cdot\w{\varphi} = 0 .
\ee
\end{itemize}
Notice that (\ref{e:divvp}) is a condition weaker than being a Killing vector
of $(\Sp_t,\w{q})$, which would write
$\mathcal{D}_\alpha \varphi_\beta + \mathcal{D}_\beta \varphi_\alpha = 0$.
For dynamical horizons, $\theta^{(\w{h})}\not=0$ and there
is a unique choice of $\w{\varphi}$ as the generator (conveniently
normalized) of the curves of constant $\theta^{(\w{h})}$
\citep{Haywa06}.

The \emph{generalized angular momentum associated with $\w{\varphi}$} is then
defined by
\be \label{e:def_J}
    J(\w{\varphi}) := - \frac{1}{8\pi} \oint_{\Sp_t}
    \langle \w{\Omega}^{(\el)}, \w{\varphi} \rangle \, \volS ,
\ee
where $\langle \w{\Omega}^{(\el)}, \w{\varphi} \rangle$ stands for the
normal fundamental form $\w{\Omega}^{(\el)}$ applied to the vector
$\w{\varphi}$ and $\volS$ is the volume element on $\Sp_t$ associated with the metric $\w{q}$.
Note that the definition of $J(\w{\varphi})$
does not depend upon the choice of null vector $\el$, thanks
to the divergence-free property of $\w{\varphi}$ and to the transformation
law (\ref{e:rescale_Omega}) of $\w{\Omega}^{(\el)}$ under a change of $\el$.
Formula (\ref{e:def_J}) coincides with \citet{AshteK03}'s definition. It also
coincides with Brown-York angular momentum \citep{BrownY93} if $\mathcal{H}$
is timelike and $\w{\varphi}$ a Killing vector.

Under the supplementary hypothesis that $\w{\varphi}$ is transported
along the evolution vector $\w{h}$ : $\Lie{h}{\w{\varphi}} = 0$,
Eq.~(\ref{e:DNS_gen}) leads to \citep{Gourg05}
\begin{eqnarray}
    & & \frac{d}{dt} J(\w{\varphi}) = - \oint_{\Sp_t} \w{T}(\w{m},\w{\varphi})
    \, \volS \nonumber \\
    & & \quad - \frac{1}{16\pi} \oint_{\Sp_t}
        \left[ \vec{\vec{\w{\sigma}}}^{(\w{m})}\!:\Lie{\varphi}\w{q}
         - 2 \theta^{(\w{k})} \w{\varphi}\cdot \DS C \right]
        \volS , \label{e:evol_J_gen}
\end{eqnarray}
where a double arrow stands for the double ``index raising'' via the
metric $\w{q}$ and the
colon denotes a double contraction. There are two interesting limiting cases
for this equation. First of all, if $\mathcal{H}$ is a null hypersurface
($C=0$ and $\w{m}=\el$), it reduces to
\begin{eqnarray}
   & & \frac{d}{dt} J(\w{\varphi}) =
    - \oint_{\Sp_t} \w{T}(\el,\w{\varphi}) \volS
     - \frac{1}{16\pi} \oint_{\Sp_t}
        \vec{\vec{\w{\sigma}}}^{(\el)}\!:\Lie{\varphi}\w{q}
        \; \volS  , \nonumber \\
	\label{e:evol_J_EH}
\end{eqnarray}
i.e. we recover Eq.~(6.134) of the \emph{Membrane Paradigm} book \citep{ThornPM86}.
Second, if $\mathcal{H}$ is a FOTH, one may show that the last term
in Eq.~(\ref{e:evol_J_gen}) vanishes \citep{Gourg05}, so that it reduces
to
\begin{eqnarray}
   \frac{d}{dt} J(\w{\varphi}) & = &
    - \oint_{\Sp_t} \w{T}(\w{m},\w{\varphi}) \volS  \nonumber \\
     & & - \frac{1}{16\pi} \oint_{\Sp_t}
        \vec{\vec{\w{\sigma}}}^{(\w{m})}\!:\Lie{\varphi}\w{q}
        \; \volS  .
\end{eqnarray}
Thus for a FOTH, the Navier-Stokes like equation (\ref{e:DNS_FOTH})
leads to an evolution equation for the angular momentum which is as
simple as Eq.~(\ref{e:evol_J_EH}) for an event horizon.
In particular the r.h.s. has only two terms, which are interpretable
as respectively (i) the flux of angular momentum due to some matter or
electromagnetic field near the horizon and (ii) the flux of angular momentum
due to ``gravitational radiation''. The latter interpretation is pretty
vague and relies on the fact that $\Lie{\varphi}\w{q} = 0$ in
axisymmetry, where gravitational radiation does not carry any angular
momentum.

\section{Area evolution and energy equation}

\subsection{Evolution of the expansion}

Let us search for an evolution equation for the
expansion $\theta^{(\w{h})}$, which governs the evolution of the
area of the surfaces $\Sp_t$ via Eq.~(\ref{e:def_expansion}).
The starting point turns out to be the Ricci identity applied to the normal vector
$\w{m}$,
as in Sec.~\ref{s:gen_non-null}, but instead of projecting it
onto $\Sp_t$ [Eq.~(\ref{e:Ricci_m_q})],
we shall project it along the normal direction to $\Sp_t$ lying in $\mathcal{H}$,
namely $\w{h}$:
\be
    \left( \nabla_\mu \nabla_\nu m^\mu
    - \nabla_\nu \nabla_\mu m^\mu \right) h^\nu  =
    R_{\mu\nu} m^\mu h^\nu .
\ee
By means of the Einstein equation, and after some computations, we arrive
at \citep{GourgJ06b}:
\begin{eqnarray}
\Lie{\w{h}} \theta^{(\w{m})} & = &  \kappa \, \theta^{(\w{h})}
	- \frac{1}{2} \theta^{(\w{h})} \theta^{(\w{m})}
	- \w{\sigma}^{(\w{h})} : \w{\sigma}^{(\w{m})}
	 \nonumber \\
	& & + \theta^{(\w{k})} \Lie{\w{h}} C
	+ \DS\cdot\left( 2 C \vec{\w{\Omega}}^{(\el)} - \vec{\DS} C \right)  \nonumber \\
	& &
	- 8\pi \w{T}(\w{m},\w{h}) , \label{e:Lh_theta_m}
\end{eqnarray}
where an upper arrow indicates ``index raising'' with the metric $\w{q}$
and the colon stands for the double contraction, i.e.
$\w{\sigma}^{(\w{h})} : \w{\sigma}^{(\w{m})} := \sigma^{(\w{h})}_{ab}
	\sigma^{(\w{m})ab}$.
If we specialize Eq.~(\ref{e:Lh_theta_m}) to the cases of (i) an event horizon
and (ii) a FOTH or a DH, we obtain respectively
\begin{eqnarray}
\Lie{\el} \theta^{(\el)} + (\theta^{(\el)})^2 - \kappa \, \theta^{(\el)} & = &
	\frac{1}{2} (\theta^{(\el)})^2
	- \w{\sigma}^{(\el)} : \w{\sigma}^{(\el)} \nonumber \\
	&&- 8\pi \w{T}(\el,\el) \label{e:evol_th_EH}
\end{eqnarray}
\begin{eqnarray}
& & \Lie{\w{h}} \theta^{(\w{h})}
	+ (\theta^{(\w{h})})^2 + \kappa \, \theta^{(\w{h})}  =
	\frac{1}{2} (\theta^{(\w{h})})^2
	+ \w{\sigma}^{(\w{h})} : \w{\sigma}^{(\w{m})} \nonumber \\
	& & \qquad - \theta^{(\w{k})} \Lie{h} C 	
	+ \DS\cdot\left( \vec{\DS} C  - 2 C \vec{\w{\Omega}}^{(\el)} \right) \nonumber \\
	& & \qquad
	+ 8\pi \w{T}(\w{m},\w{h}) .  \label{e:evol_th_FOTH}
\end{eqnarray}
For the event horizon, we have used the null character of $\mathcal{H}$, which implies $C=0$ and $\w{h}=\w{m}=\el$, yielding Eq.~(\ref{e:evol_th_EH}). It is nothing but
the null Raychaudhuri equation for a surface-orthogonal congruence
\citep{HawkiH72}.
For the FOTH/DH case [Eq.~(\ref{e:evol_th_FOTH})], we have used the property $\theta^{(\w{m})}=-\theta^{(\w{h})}$ already encountered in Sec.~\ref{s:gen_non-null}.
Notice the change of some signs between Eqs.~(\ref{e:evol_th_EH}) and (\ref{e:evol_th_FOTH}).

\subsection{Energy dissipation and bulk viscosity}

In the fluid membrane approach to black holes, \citet{PriceT86} and \citet{ThornPM86} defined the \emph{surface energy
density} of an event horizon as $\varepsilon := -\theta^{(\el)}/8\pi$ and
interpreted Eq.~(\ref{e:evol_th_EH}) as an energy balance law,
with heat production resulting from viscous stresses.
By analogy, let us define the \emph{surface energy density} of a FOTH/DH
as $\varepsilon := -\theta^{(\w{m})}/8\pi$, where the role of the normal to $\mathcal{H}$
is now taken by $\w{m}$ instead of $\el$. Since
$\theta^{(\w{m})} = -\theta^{(\w{h})}$ for a FOTH/DH, we have
\be
\varepsilon = \frac{\theta^{(\w{h})}}{8\pi}
\ee
and we may rewrite Eq.~(\ref{e:evol_th_FOTH}) as
\begin{eqnarray}
	\Lie{h}\varepsilon + \theta^{(\w{h})} \varepsilon & = &
	- \frac{\kappa}{8\pi} \theta^{(\w{h})}
	+  \frac{1}{8\pi} \w{\sigma}^{(\w{h})} : \w{\sigma}^{(\w{m})}
	+ \frac{(\theta^{(\w{h})})^2}{16\pi}   \nonumber \\
	& &
	- \DS\cdot\w{Q}
	+ \w{T}(\w{m},\w{h})
	- \frac{\theta^{(\w{k})}}{8\pi} \Lie{h} C ,  \label{e:cons_ener}
\end{eqnarray}
with $\w{Q} := \frac{1}{4\pi} \left[ C \vec{\w{\Omega}}^{(\el)} - 1/2\, \vec{\DS} C  \right] = - \frac{C}{4\pi} \vec{\w{\varpi}}$, where $\w{\varpi}$ is the
\emph{anholonomicity 1-form} (or \emph{twist 1-form}) of the 2-surface
$\Sp_t$ \citep{Haywa94} \citep[see also Sec.~IV.A of][]{Gourg05} and
$\vec{\w{\varpi}}$ denotes its vector dual.

It is striking that Eqs.~(\ref{e:DNS_FOTH}) and (\ref{e:cons_ener}) are fully analogous to
the equations that govern a two-dimensional
non-relativistic fluid of internal energy density $\varepsilon$, momentum density
$\w{\pi}$, pressure $\kappa/8\pi$,
shear stress tensor $\w{\sigma}^{(\w{m})}/8\pi$,
bulk viscosity $\zeta = 1/16\pi$, shear strain tensor $\w{\sigma}^{(\w{h})}$, expansion $\theta^{(\w{h})}$, subject to the external force density
$- \w{T}(\w{m},\vec{\w{q}}) + \theta^{(\w{k})}/8\pi\,  \DS C$, external energy production rate
$\w{T}(\w{m},\w{h})- \theta^{(\w{k})}/8\pi\, \Lie{h} C $ and  heat flux $\w{Q}$
\citep[see e.g.][]{Rieut97}.
In particular the value of the bulk viscosity read on Eq.~(\ref{e:cons_ener})
(the coefficient of $(\theta^{(\w{h})})^2$) is the same as that obtained
from the Navier-Stokes equation (\ref{e:DNS_FOTH}) and given by
Eq.~(\ref{e:zeta_FOTH}).

Besides, let us notice that the shear viscosity $\mu$ does not appear in
Eqs.~(\ref{e:DNS_FOTH}) and (\ref{e:cons_ener}), because the standard
Newtonian-fluid relation between the shear stress tensor $\w{\sigma}^{(\w{m})}/8\pi$ and the shear strain tensor $\w{\sigma}^{(\w{h})}$, namely $\w{\sigma}^{(\w{m})}/8\pi = 2 \mu \w{\sigma}^{(\w{h})}$, does not hold.
Here we have
$\w{\sigma}^{(\w{m})}/8\pi =
[ \w{\sigma}^{(\w{h})} + 2 C \w{\sigma}^{(\w{k})}]/8\pi$, so that the
Newtonian-fluid assumption is fulfilled only if $C=0$ (isolated horizon limit).
On the contrary, it appears from Eqs.~(\ref{e:DNS_FOTH}) and (\ref{e:cons_ener}) that
the trace part of the viscous stress tensor $\w{S}_{\rm visc}$ does obey the
Newtonian-fluid law, being proportional to the
trace part of the strain tensor (i.e. the expansion $\theta^{(\w{h})}$):
${\rm tr} \, \w{S}_{\rm visc} = 3 \zeta \theta^{(\w{h})}$.

We may point out two differences with the event horizon case
\citep{Damou79,Damou82,PriceT86,ThornPM86}. First the heat flux $\w{Q}$ is not
vanishing for a FOTH/DH, whereas it was zero for an EH.
Notice that $\w{Q}$ is a vector tangent to $\Sp_t$ so that the integration
of Eq.~(\ref{e:cons_ener}) over the closed surface $\Sp_t$ to get a global
internal energy balance law would not contain any net heat flux.
The second major difference is that, as already stressed in
Sec.~\ref{s:gen_non-null}, the bulk viscosity
$\zeta$ is positive, being equal to $1/16\pi$ [Eq.~(\ref{e:zeta_FOTH})],
whereas it was found to be negative, being equal to $-1/16\pi$ [Eq.~(\ref{e:zeta_EH})],
for an event horizon. As commented in Sec.~\ref{s:DNS_ori}, this
negative value is related to the teleological character of event horizons.
On the contrary the positive value of the bulk viscosity for FOTHs and DHs that
these objects behave as ``ordinary'' physical objects and is in perfect agreement
with their local nature.

\section*{Acknowledgements}
EG would like to express his deep gratitude to Jean-Pierre Lasota, who as
a professor in a master course in Meudon, initiated him to the wonderful field of relativistic astrophysics and black holes. Since then, Jean-Pierre has been of unfailing support and it is a great pleasure to dedicate this paper to him.
We would also like to thank greatly Marek Abramowicz for having organized the splendid
Trzebieszowice conference in honour of Jean-Pierre.
This work was supported by the ANR grant 06-2-134423 \emph{M\'ethodes math\'ematiques pour la relativit\'e g\'en\'erale}.

\bibliographystyle{elsart-harv}

\begin{thebibliography}{36}
\expandafter\ifx\csname natexlab\endcsname\relax\def\natexlab#1{#1}\fi


\bibitem[Andersson \etal(2005)]{AnderMS05} Andersson L., Mars M. \&  Simon W., 2005, {\em  Local Existence of Dynamical and Trapping Horizons}, Phys. Rev. Lett. {\bf 95} 111102 

\bibitem[Andersson \& Metzger(2008)]{AnderM07}
Andersson L. \& Metzger J., 2008, {\em  The area of horizons and the trapped region}, preprint arXiv:0708.4252.

\bibitem[Ansorg(2005)]{Ans05}
Ansorg M., 2005,
{\em A double-domain spectral method for black hole excision data},
Phys. Rev. D {\bf 72}, 024018

\bibitem[Ashtekar \etal(1999)]{AshteBF99}
Ashtekar A., Beetle C. \& Fairhurst S., 1999,
{\em Isolated horizons: a generalization of black hole mechanics},
Class. Quantum Grav. {\bf 16}, L1

\bibitem[Ashtekar \& Galloway(2005)]{AshteG05}
Ashtekar A. \& Galloway G.J., 2005,
{\em Some uniqueness results for dynamical horizons},
Adv. Theor. Math. Phys. {\bf 9}, 1

\bibitem[Ashtekar \& Krishnan(2002)]{AshteK02}
Ashtekar A. \& Krishnan B., 2002,
{\em Dynamical Horizons: Energy, Angular Momentum, Fluxes and Balance Laws},
Phys. Rev. Lett. {\bf 89} 261101

\bibitem[Ashtekar \& Krishnan(2003)]{AshteK03}
Ashtekar A. \& Krishnan B., 2003,
{\em Dynamical horizons and their properties},
Phys. Rev. D {\bf 68}, 104030

\bibitem[Ashtekar \& Krishnan(2004)]{AshteK04}
Ashtekar A. \& Krishnan B., 2004,
{\em Isolated and dynamical horizons and their applications},
Living Rev. Relativity {\bf 7}, 10 [Online article],
{\tt http://www.livingreviews.org/lrr-2004-10}

\bibitem[Booth(2005)]{Booth05}
Booth I, 2005,
{\em Black hole boundaries},
Canadian J. Phys. {\bf 83}, 1073

\bibitem[Booth \& Fairhurst(2004)]{BoothF04}
Booth I. \& Fairhurst S., 2004,
{\em The first law for slowly evolving horizons},
Phys. Rev. Lett. {\bf 92}, 011102

\bibitem[Booth \& Fairhurst(2005)]{BoothF05}
Booth I. \& Fairhurst S., 2005,
{\em Horizon energy and angular momentum from a Hamiltonian perspective},
Class. Quantum Grav. {\bf 22}, 4545

\bibitem[Brown \& York(1993)]{BrownY93}
Brown J. D. \&  York J. W., 1993,
{\em  Quasilocal energy and conserved charges derived from the gravitational action},
Phys. Rev. D {\bf 47}, 1407

\bibitem[Carter(1992)]{Carte92a}
Carter B., 1992,
{\em Outer curvature and conformal geometry of an imbedding},
J. Geom. Phys. {\bf 8}, 53

\bibitem[Caudill \etal(2006)]{CauCooGri06}
Caudill M., Cook G.~B., Grigsby  J.~D. \& Pfeiffer H.~P., 2006,
{\em Circular orbits and spin in black-hole initial data},
Phys. Rev. D {\bf 74}, 064011

\bibitem[Cook \& Pfeiffer(2004)]{CooPfe04}
Cook G.~B. \& Pfeiffer H.~P., 2004,
{\em Excision boundary conditions for black-hole initial data},
Phys. Rev. D {\bf 70}, 104016

\bibitem[Dain \etal(2005)]{DaiJarKri05}
Dain S., Jaramillo J.~L. \& Krishnan B., 2005,
{\em On the existence of initial data containing isolated black holes},
Phys. Rev. D {\bf 71}, 064003

\bibitem[Damour(1979)]{Damou79}
Damour T., 1979,
{\em Quelques propri\'et\'es m\'ecaniques, \'electromagn\'etiques, thermo\-dy\-na\-mi\-ques et quantiques des trous noirs}, Th\`ese de doctorat d'\'Etat, Universit\'e Paris 6

\bibitem[Damour(1982)]{Damou82}
Damour T., 1982,
{\em Surface effects in black hole physics},
in {\em Proceedings of the Second Marcel Grossmann Meeting on General
Relativity}, Ed. R.~Ruffini, North Holland, p.~587

\bibitem[Damour \& Lilley(2008)]{DamouL08}
Damour T. \& Lilley M., 2008,
{\em String theory, gravity and experiment},
lectures at Les Houches Summer School in Theoretical Physics (2-27 July 2007);
preprint arXiv:0802.4169.

\bibitem[Demia\'nski \& Lasota(1973)]{DemiaL73}
Demia\'nski M. \& Lasota J.-P., 1973,
{\em Black holes in an expanding Universe},
Nature Phys. Sci. {\bf 241}, 53

\bibitem[Gourgoulhon(2005)]{Gourg05}
Gourgoulhon E., 2005,
{\em Generalized Damour-Navier-Stokes equation applied to
trapping horizons}, Phys. Rev. D {\bf 72}, 104007

\bibitem[Gourgoulhon \& Jaramillo(2006a)]{GourgJ06a}
Gourgoulhon E. \& Jaramillo J.L., 2006a, 
{\em A 3+1 perspective on null hypersurfaces and isolated horizons},
Phys. Rep. {\bf 423}, 159

\bibitem[Gourgoulhon \& Jaramillo(2006b)]{GourgJ06b}
Gourgoulhon E. \&  Jaramillo J.L., 2006b,
{\em  Area evolution, bulk viscosity, and entropy principles for dynamical horizons},
Phys. Rev. D {\bf 74}, 087502

\bibitem[H\'a\'\j i\v{c}ek(1973)]{Hajic73}
H\'a\'\j i\v{c}ek P., 1973,
{\em Exact models of charged black holes. I. Geometry of totally geodesic null hypersurface},
Commun. Math. Phys. {\bf 34}, 37

\bibitem[Hartle(1973)]{Hartl73}
Hartle  J.B., 1973,
{\em Tidal friction in slowly rotating black holes},
Phys. Rev. D {\bf 8}, 1010

\bibitem[Hawking(1972)]{Hawki72}
Hawking S.W., 1972,
{\em Black holes in general relativity},
Commun. Math. Phys. {\bf 25}, 152

\bibitem[Hawking \&  Ellis(1973)]{HawkiE73}
Hawking S.W. \&  Ellis G.F.R., 1973,
{\em The large scale structure of
space-time},
Cambridge University Press, Cambridge

\bibitem[Hawking \& Hartle(1972)]{HawkiH72}
Hawking S.W. \& Hartle  J.B., 1972,
{\em Energy and angular momentum flow into a black hole},
Commun. Math. Phys. {\bf 27}, 283

\bibitem[Hayward(1994)]{Haywa94}
Hayward S.A., 1994,
{\em General laws of black hole dynamics},
Phys. Rev. D {\bf 49}, 6467

\bibitem[Hayward(2004)]{Haywa04b}
Hayward S.A., 2004,
{\em Energy and entropy conservation for dynamical black holes},
Phys. Rev. D {\bf 70}, 104027

\bibitem[Hayward(2006)]{Haywa06}
Hayward S.A., 2006,
{\em Angular momentum conservation for dynamical black holes},
Phys.Rev. D {\bf 74}, 104013

\bibitem[Jaramillo \etal(2007)]{JaramAL07}
Jaramillo J.L., Ansorg M. \& Limousin F., 2007,
{\em Numerical implementation of isolated horizon boundary conditions},
Phys.Rev. D {\bf 75}, 024019

\bibitem[Jaramillo \etal(2008a)]{JaramGCI08}
Jaramillo J.L., Gourgoulhon E., Cordero-Carri\'on I. \&  Ib\'a{\~n}ez J.M.,
2008a, {\em Trapping Horizons as inner boundary conditions for black hole spacetimes}, 
Phys. Rev. D {\bf 77}, 047501

\bibitem[Jaramillo \etal(2004)]{JarGouMen04}
Jaramillo J.~L., Gourgoulhon E. \&  Mena~Marug\'an G.~A., 2004,
{\em Inner boundary conditions for black hole initial data derived from isolated horizons},
Phys. Rev. D {\bf 70}, 124036

\bibitem[Jaramillo \etal(2008b)]{JaramVG08}
Jaramillo J.L., Valiente Kroon J.A. \& Gourgoulhon E., 2008b,
{\em From Geometry to Numerics: interdisciplinary aspects in mathematical and numerical relativity}, Class. Quantum Grav., in press, preprint arXiv:0712.2332

\bibitem[Krishnan(2008)]{Krish08}
Krishnan B., 2008, {\em Fundamental properties and applications of
quasi-local black hole horizons},
to appear in the Proceedings of the 18th International Conference on General Relativity and Gravitation (Sidney, Australia, 8-13 July 2007);
preprint arXiv:0712.1575

\bibitem[Penrose(1965)]{Penro65}
Penrose R., 1965, {\em Gravitational collapse and space-time singularities},
Phys. Rev. Lett. {\bf 14}, 57

\bibitem[Price \& Thorne(1986)]{PriceT86}
Price R.H. \& Thorne K.S., 1986, {\em Membrane viewpoint on black holes:
Properties and evolution of the stretched horizon},
Phys. Rev. D {\bf 33}, 915

\bibitem[Rieutord(1997)]{Rieut97}
Rieutord M., 1997, \emph{Une introduction \`a la dynamique des fluides},
Masson, Paris

\bibitem[Schnetter \etal(2006)]{SchneKB06}
Schnetter E., Krishnan B. \& Beyer F., 2006,
{\em  Introduction to dynamical horizons in numerical relativity},
Phys.Rev. D {\bf 74}, 024028

\bibitem[Senovilla(2005)]{Senov05}
Senovilla J.M.M., 2005, {\em Trapped submanifolds in Lorentzian geometry},
in Proc. XIII Fall Workshop on Geometry and Physics
(Murcia, Spain, 2004), edited by L.J. Al\'\i as, A. Ferr\'andez,
M.A. Hen\'andez, P. Lusca, and J.A. Pastor,
Public. Real Soc. Mat. Esp. {\bf 9}; also available as
arXiv:math.DG/0412256

\bibitem[Thorne \etal(1986)]{ThornPM86}
Thorne K.S., Price R.H. \& MacDonald D.A., 1986, {\em Black holes : the membrane paradigm},
Yale University Press, New Haven

\end{thebibliography}

\end{document}